\documentclass{revtex4-1}

\usepackage{amssymb,amsmath,xcolor,graphicx,xspace,colortbl,textcomp, rotating} 
\usepackage{amssymb,amsmath,xcolor,graphicx,xspace,colortbl,textcomp, rotating}  
\usepackage{endnotes}  
\usepackage{textcomp}  
\graphicspath{{E_and_C_gravity_final draft v2_graphics/}{E_and_C_gravity_final draft v2_tcache/}{E_and_C_gravity_final draft v2_gcache/}}
\DeclareGraphicsExtensions{.pdf,.eps,.ps,.png,.jpg,.jpeg}
\graphicspath{{E_and_C_gravity final draft_graphics/}{E_and_C_gravity final draft_tcache/}{E_and_C_gravity final draft_gcache/}}
\DeclareGraphicsExtensions{.pdf,.eps,.ps,.png,.jpg,.jpeg}
\graphicspath{{E_and_C_gravity draft08_graphics/}{E_and_C_gravity draft08_tcache/}{E_and_C_gravity draft08_gcache/}}
\DeclareGraphicsExtensions{.pdf,.eps,.ps,.png,.jpg,.jpeg}
\graphicspath{{E_and_C_gravity draft08_graphics/}{E_and_C_gravity draft08_tcache/}{E_and_C_gravity draft08_gcache/}}
\DeclareGraphicsExtensions{.pdf,.eps,.ps,.png,.jpg,.jpeg}
\graphicspath{{E_and_C_gravity draft08_graphics/}{E_and_C_gravity draft08_tcache/}{E_and_C_gravity draft08_gcache/}}
\DeclareGraphicsExtensions{.pdf,.eps,.ps,.png,.jpg,.jpeg}
\graphicspath{{E_and_C_gravity draft07_graphics/}{E_and_C_gravity draft07_tcache/}{E_and_C_gravity
draft07_gcache/}} \DeclareGraphicsExtensions{.pdf,.eps,.ps,.png,.jpg,.jpeg}
\graphicspath{{E_and_C_gravity draft07_graphics/}{E_and_C_gravity draft07_tcache/}{E_and_C_gravity
draft07_gcache/}} \DeclareGraphicsExtensions{.pdf,.eps,.ps,.png,.jpg,.jpeg}
\graphicspath{{E_and_C_gravity draft07_graphics/}{E_and_C_gravity draft07_tcache/}{E_and_C_gravity
draft07_gcache/}} \DeclareGraphicsExtensions{.pdf,.eps,.ps,.png,.jpg,.jpeg}
\graphicspath{{E_and_C gravity draft06_graphics/}{E_and_C gravity draft06_tcache/}{E_and_C gravity
draft06_gcache/}} \DeclareGraphicsExtensions{.pdf,.eps,.ps,.png,.jpg,.jpeg}
\graphicspath{{E_and_C gravity draft06_graphics/}{E_and_C gravity draft06_tcache/}{E_and_C gravity
draft06_gcache/}} \DeclareGraphicsExtensions{.pdf,.eps,.ps,.png,.jpg,.jpeg}
\graphicspath{{E_and_C gravity draft06_graphics/}{E_and_C gravity draft06_tcache/}{E_and_C gravity
draft06_gcache/}} \DeclareGraphicsExtensions{.pdf,.eps,.ps,.png,.jpg,.jpeg}
\graphicspath{{E_and C gravity draft04_graphics/}{E_and C gravity draft04_tcache/}{E_and C gravity
draft04_gcache/}} \DeclareGraphicsExtensions{.pdf,.eps,.ps,.png,.jpg,.jpeg}
\graphicspath{{E_and C gravity draft04_graphics/}{E_and C gravity draft04_tcache/}{E_and C gravity
draft04_gcache/}} \DeclareGraphicsExtensions{.pdf,.eps,.ps,.png,.jpg,.jpeg}
\graphicspath{{E_and C gravity draft04_graphics/}{E_and C gravity draft04_tcache/}{E_and C gravity
draft04_gcache/}} \DeclareGraphicsExtensions{.pdf,.eps,.ps,.png,.jpg,.jpeg}
\graphicspath{{E_and C gravity draft03_graphics/}{E_and C gravity draft03_tcache/}{E_and C gravity
draft03_gcache/}} \DeclareGraphicsExtensions{.pdf,.eps,.ps,.png,.jpg,.jpeg}

\input{tcilatex}
\begin{document}
\title{From conformal to Einstein gravity }
\author{Giorgos Anastasiou }
\email{georgios.anastasiou@unab.cl }
\author{Rodrigo Olea }
\email{ rodrigo.olea@unab.cl }
\affiliation{Departamento de Ciencias F{\'\i}sicas, Universidad Andres Bello, Sazi{\'e} 2212, Piso 7, Santiago, Chile }
\begin{abstract}We provide a simple derivation of the equivalence between Einstein and conformal gravity (CG) with Neumann boundary conditions given by Maldacena.
As Einstein spacetimes are Bach flat, a generic solution to CG would contain both Einstein and non-Einstein parts. Using this decomposition of the spacetime
curvature in the Weyl tensor makes manifest the equivalence between the two theories, both at the level of the action and the variation of it. As a consequence,
we show that the on-shell action for critical gravity in four dimensions is given uniquely in terms of the Bach tensor. 
\end{abstract}
\maketitle

\section{Introduction}
In the last decades, modifications to general relativity (GR) have been extensively studied in a quest for a quantum theory of gravity. In this
context, GR can be considered as a low-energy effective theory of gravity that may acquire higher-order corrections. The presence of nonlinear terms in
curvature exhibits desirable features one may expect from quantum gravity: unlike GR, which is a two-loop divergent theory \cite{Goroff1985},
higher-derivative gravity theories are renormalizable providing a well-defined ultraviolet behavior \cite{Stelle1977,Duff1975}.
Examples of this type of modification are massive gravity, $F \left (R\right )$, etc. 

Conformal gravity (CG) in four dimensions is a special case of the aforementioned class of theories. It
is a quadratic-curvature gravity theory that is renormalizable \cite{Adler1982}
but it suffers from ghosts (modes of negative norm), contrary to Einstein (E) gravity. Its action takes the form
\begin{equation}I_{CG} =\alpha _{CG}\int _{M}d^{4}x\sqrt{ -g} W^{\alpha  \beta  \mu  \nu } W_{\alpha  \beta  \mu  \nu } =\frac{\alpha _{CG}}{4}\int _{M}d^{4}x\sqrt{ -g} \delta _{\left [\mu _{1} \mu _{2} \mu _{3} \mu _{4}\right ]}^{\left [\nu _{1} \nu _{2} \nu _{3} \nu _{4}\right ]}W_{\nu _{1}\nu _{2}}^{\mu _{1}\mu _{2}}W_{\nu _{3}\nu _{4}}^{\mu _{3}\mu _{4}}\thinspace  , \label{confgraction}
\end{equation}  where \begin{equation}W_{\mu \nu }^{\alpha \beta } =R_{\mu \nu }^{\alpha \beta } -(S_{\mu }^{\alpha } \delta _{\nu }^{\beta } -S_{\mu }^{\beta } \delta _{\nu }^{\alpha } -S_{\nu }^{\alpha } \delta _{\mu }^{\beta } +S_{\nu }^{\beta } \delta _{\mu }^{\alpha })\thinspace  ,
\end{equation} is the Weyl tensor in terms of the Schouten tensor\begin{equation}S_{\mu }^{\alpha } =\frac{1}{2}(R_{\mu }^{\alpha } -\frac{1}{6}R\delta _{\mu }^{\alpha }) , \label{schouten}
\end{equation} and $\delta _{\left [\cdots \right ]}^{\left [\cdots \right ]}$ is the totally antisymmetric product of Kronecker deltas. 

This theory is invariant under local Weyl rescalings of the metric
$g_{\mu  \nu } \rightarrow \Omega ^{2} \left (x\right ) g_{\mu  \nu }\thinspace  ,$ i.e., a transformation that preserves the angles but not the distances. The coupling constant in front of CG action, $\alpha _{C G}$ , is a positive dimensionless parameter. At a more fundamental level, the Lagrangian for CG can be obtained from twistor string theory
\cite{Berkovits2004} .

Furthermore, Maldacena has
shown recently an interesting connection between Einstein gravity with cosmological constant and CG in four dimensions \cite{Maldacena2011}
: at the tree level, four-dimensional CG with Neumann boundary conditions is equivalent to Einstein gravity with cosmological constant. 

In
a sense, this is the converse argument to the one given in a prior work \cite{Miskovic2009}:
the regularized action for anti-de Sitter (AdS) gravity is on-shell equivalent to the action of CG for a given coupling constant $\alpha _{C G} =\ell ^{2}/64 \pi  G$ , where $\ell $ is the AdS radius and $G$ is the Newton constant. This proof relies on the fact that the boundary counterterms that render the AdS action and its variation
finite, as prescribed in the context of gauge/gravity duality \cite{deHaro2000},
can be summed up in the addition of a single topological invariant in the bulk \cite{Miskovic2009,Miskovic2014}.

While the latter argument uses Einstein spaces in AdS gravity, CG in four dimensions possesses a broader class of solutions of which
Einstein spaces are only a particular subset. With this in mind, in CG the Weyl tensor splits in two parts: an Einstein part linked to the Einstein
spacetimes and a non-Einstein (NE) part that is linear on the Bach tensor. 

In this paper, the decomposition mentioned above is crucial
to isolate the Einstein part of the CG action from the higher-derivative contributions coming from the Bach tensor. In particular, this leads to a simpler
proof of the equivalence between CG and Einstein gravity given in Ref. \cite{Maldacena2011}.

\section{Conformal Gravity: Field Equations and Surface Terms}
Having already defined the CG action in Eq.(\ref{confgraction}) we turn to the first variation
of it. After some algebraic manipulation, the variation of the action can be written as
\begin{multline}\delta  I_{C G} =\alpha _{C G} \int _{M}d^{4} x \sqrt{ -g} B_{\mu }^{\nu } \left (g^{ -1} \delta  g\right )_{\nu }^{\mu } + \\
 +\alpha _{C G} \int _{ \partial M}d^{3} x \sqrt{ -h} \delta _{\left [\mu _{1} \mu _{2} \mu _{3} \mu _{4}\right ]}^{\left [\nu _{1} \nu _{2} \nu _{3} \nu _{4}\right ]} \left [n_{\nu _{1}} \delta  \Gamma _{\kappa  \nu _{2}}^{\mu _{1}} g^{\mu _{2} \kappa } W_{\nu _{3} \nu _{4}}^{\mu _{3} \mu _{4}} +n^{\mu _{1}}  \nabla _{\nu _{1}}W_{\nu _{2} \nu _{3}}^{\mu _{2} \mu _{3}} \left (g^{ -1} \delta  g\right )_{\nu _{4}}^{\mu _{4}}\right ] , \label{confgravar}\end{multline}  where $n_{\mu }$ is normal to the boundary and the term in the bulk is the Bach tensor,
\begin{equation}B_{\mu }^{\nu } = -\delta _{\left [\mu  \mu _{1} \mu _{2} \mu _{3}\right ]}^{\left [\nu  \nu _{1} \nu _{2} \nu _{3}\right ]} \left [ \nabla ^{\mu _{1}} \nabla _{\nu _{1}}W_{\nu _{2} \nu _{3}}^{\mu _{2} \mu _{3}} +\frac{1}{2} R_{\nu _{1}}^{\mu _{1}} W_{\nu _{2} \nu _{3}}^{\mu _{2} \mu _{3}}\right ] = -4 \left [ \nabla ^{\alpha } \nabla _{\beta }W_{\alpha  \mu }^{\beta  \nu } +\frac{1}{2} R_{\beta }^{\alpha } W_{\alpha  \mu }^{\beta  \nu }\right ] ,
\end{equation}  which is four derivative, traceless, and covariantly conserved. Therefore, the equations of motion (EOM) of CG are satisfied by Bach flat
solutions. Einstein spacetimes are Bach-flat and constitute a trivial subset of CG solutions. 

When added on top of the Einstein gravity action with negative cosmological constant $\Lambda  = -3/\ell ^{2}$ ,
\begin{equation}I^{(E)} =\frac{1}{16\pi G}\int _{M}d^{4}x \sqrt{ -g}(R -2\Lambda ) ,
\end{equation}the EOM is $G_{\mu }^{\nu } -\gamma B_{\mu }^{\nu } =0$ where $G_{\mu }^{\nu }$ is the Einstein tensor and $\gamma $  is given in terms of the CG coupling as $\gamma  =16\pi G\alpha _{CG}$. Taking the trace of this relation,
the Ricci tensor is
\begin{equation}R_{\mu  \nu } = -\frac{3}{\ell ^{2}}g_{\mu \nu } +\gamma  B_{\mu  \nu }\; , \label{RicciEOM}
\end{equation}   which is the most general form in Einstein-Weyl gravity. Considering arbitrary quadratic couplings in the curvature would have modified the asymptotic behavior such that a unique AdS vacuum would no longer exist.

The
departure from second-order field equations in CG switches on both massless and massive modes, which appear to have opposite norms. The massive modes are
also characterized by partial masslessness \cite{Deser1983,Deser2001}, i.e.,
the scalar component of the massive mode is absent and the only remaining parts of the decomposition are the ones corresponding to spin 2 and the spin 1.
The appearance of ghosts is the price one has to pay in order to render the theory renormalizable. Recent work \cite{Bender2007}
has shown that there is a specific realization of the theory that is unitary and where, indeed, no ghost modes arise. 

\section{Topological Regularization in Einstein AdS gravity}
The action for Einstein gravity with negative cosmological constant is proportional to the volume of asymptotically AdS spacetimes, which is infinite. In the context of gauge/gravity
duality, the problem of extracting the physical information of the field theory residing at the boundary of the spacetime reduces to the problem of volume
renormalization. Holographic renormalization provides a prescription that renders the on-shell action finite by the addition of local counterterms \cite{deHaro2000}.

It was shown in Ref. \cite{Miskovic2009} that the
addition to the Einstein AdS gravity action of the Gauss-Bonnet (GB) term in four dimensions, i.e.,
\begin{equation}I_{ren}^{\left (E\right )} =\frac{1}{16 \pi  G} \int _{M}d^{4} x \sqrt{ -g} \left [R -2 \Lambda  +\frac{\ell ^{2}}{16} \delta _{\left [\mu _{1} \mu _{2} \mu _{3} \mu _{4}\right ]}^{\left [\nu _{1} \nu _{2} \nu _{3} \nu _{4}\right ]} R_{\nu _{1} \nu _{2}}^{\mu _{1} \mu _{2}} R_{\nu _{3} \nu _{4}}^{\mu _{3} \mu _{4}}\right ]\;\;\text{,}\; \label{adsaction}
\end{equation}
and, in general, the addition of the Euler term in higher even dimensions, is equivalent to the program of holographic renormalization, as boundary terms
are concerned. Though it may at first sound surprising, from the mathematical viewpoint, this result is justified by the fact that the GB term can be written
as a surface term that depends both on intrinsic and extrinsic curvatures of the boundary ( $K_{j}^{i}$ and $\mathcal{R}_{j l}^{i k}$ , respectively) \cite{Olea2005},
\begin{equation}I_{ren}^{\left (E\right )} =\frac{1}{16 \pi  G} \int _{M}d^{4} x \sqrt{ -g} \left (R -2 \Lambda \right ) +\frac{\ell ^{2}}{16 \pi  G} \int _{ \partial M}d^{3} x \sqrt{ -h} \delta _{\left [i_{1} i_{2} i_{3}\right ]}^{\left [j_{1} j_{2} j_{3}\right ]} K_{j_{1}}^{i_{1}} \left (\frac{1}{2} \mathcal{R}_{j_{2} j_{3}}^{i_{2} i_{3}} -\frac{1}{3} K_{j_{2}}^{i_{2}} K_{j_{3}}^{i_{3}}\right )\;\;\text{,}\;\text{\ \ }
\end{equation}  up to the Euler characteristic of the manifold $\chi  (M)$ . Here, $h_{i j}$ is the boundary three-dimensional metric. The polynomial in $K_{j}^{i}$ and $\mathcal{R}_{j l}^{i k}$ acts as a series of \emph{extrinsic} counterterms. 

The asymptotic resolution of the Einstein
equations allows us to write down an expansion of the extrinsic curvature in terms of intrinsic quantities of the boundary. This is the key ingredient to generate
the standard counterterms in Ref. \cite{Balasubramanian2000} from a topological
invariant \cite{Miskovic2009}. This argument provides a source of certainty
that we can properly refer to $I_{ren}^{\left (E\right )}$ as the renormalized action obtained by holographic techniques in asymptotically AdS gravity. 

Because of the addition of a bulk
topological invariant instead of boundary counterterms, the renormalized action features a remarkable property, which is
\begin{equation}I_{r e n}^{\left (E\right )} =\frac{\ell ^{2}}{256 \pi  G} \int _{M}d^{4} x \sqrt{ -g} \delta _{\left [\mu _{1} \mu _{2} \mu _{3} \mu _{4}\right ]}^{\left [\nu _{1} \nu _{2} \nu _{3} \nu _{4}\right ]} W_{\left (E\right ) \nu _{1} \nu _{2}}^{\mu _{1} \mu _{2}} W_{\left (E\right ) \nu _{3} \nu _{4}}^{\mu _{3} \mu _{4}}\; , \label{renadsaction}
\end{equation}  where
\begin{equation}W_{\left (E\right ) \nu _{1} \nu _{2}}^{\mu _{1} \mu _{2}} =R_{\nu _{1} \nu _{2}}^{\mu _{1} \mu _{2}} +\frac{1}{\ell ^{2}} \delta _{[\nu _{1} \nu _{2}]}^{[\mu _{1} \mu _{2}]} \label{WeylE}
\end{equation}  is the Weyl tensor for any Einstein space ( $R_{\mu  \nu } = -(3/\ell ^{2}) g_{\mu  \nu }$ ) \cite{commAnders}. 

Furthermore,
taking variations of Eq.(\ref{adsaction}), one obtains
\begin{gather}\delta  I_{r e n}^{\left (E\right )} =\frac{\ell ^{2}}{64 \pi  G} \int _{ \partial M}d^{3} x \sqrt{ -h} \delta _{\left [\mu _{1} \mu _{2} \mu _{3} \mu _{4}\right ]}^{\left [\nu _{1} \nu _{2} \nu _{3} \nu _{4}\right ]} n_{\nu _{1}} \delta  \Gamma _{\kappa  \nu _{2}}^{\mu _{1}} g^{\mu _{2} }(R_{\nu _{3}\nu _{4}}^{\mu _{3}\mu _{4}} +\frac{1}{\ell ^{2}}\delta _{[\nu _{3}\nu _{4}]}^{[\mu _{3}\mu _{4}]}) \nonumber  \\
 =\frac{\ell ^{2}}{64 \pi  G} \int _{ \partial M}d^{3}x \sqrt{ -h} \delta _{\left [\mu _{1} \mu _{2} \mu _{3} \mu _{4}\right ]}^{\left [\nu _{1} \nu _{2} \nu _{3} \nu _{4}\right ]} n_{\nu _{1}} \delta  \Gamma _{\kappa  \nu _{2}}^{\mu _{1}} g^{\mu _{2} \kappa } W_{(E)\nu _{3}\nu _{4}}^{\mu _{3}\mu _{4}} \label{varadsaction}\end{gather}  when the EOM hold. We have used Eq.(\ref{WeylE}) and the fact that the GB term does not contribute to the EOM. The form adopted by the renormalized action for Einstein AdS gravity (\ref{renadsaction})
and its corresponding variation (\ref{varadsaction}) plays a significant role in the analysis in the next
section, where we isolate the Einstein part in the action for CG. 

\section{From Conformal to Einstein Gravity}
Thinking of a general solution to CG as a deviation from the Einstein branch, our starting point is Eq.(\ref{RicciEOM}) in order to make a connection to the theory in the previous section. In this formulation, the Bach tensor represents a wider
class of solutions and captures the higher-derivative contributions in the action in CG, such that we separate the Weyl tensor $W =W_{(E)} +W_{(N E)}$ into an E and a NE part. For the Schouten tensor (\ref{schouten}), one gets
\begin{equation}S_{\mu }^{\alpha } = -\frac{1}{2} \left (\frac{1}{\ell ^{2}} \delta _{\mu }^{\alpha } -\gamma  B_{\mu }^{\alpha }\right )\text{\thinspace \ }\;\text{,}\;\;
\end{equation}  such that the Einstein part of the Weyl tensor matches Eq.(\ref{WeylE}).
In turn, the non-Einstein part is a skew-symmetric product of the Bach tensor and the metric
\begin{equation}W_{\left (N E\right ) \mu \nu }^{\alpha \beta } = -\frac{\gamma }{2} (B_{\mu }^{\alpha }\delta _{\nu }^{\beta } -B_{\mu }^{\beta } \delta _{\nu }^{\alpha } -B_{\nu }^{\alpha } \delta _{\mu }^{\beta } +B_{\nu }^{\beta } \delta _{\mu }^{\alpha })\;\;\text{.}\;\;
\end{equation}

Hence, the CG action can be rewritten as
\begin{equation}I_{C G} =\frac{\alpha _{C G}}{4} \int _{M}d^{4} x \sqrt{ -g} \delta _{\left [\mu _{1} \mu _{2} \mu _{3} \mu _{4}\right ]}^{\left [\nu _{1} \nu _{2} \nu _{3} \nu _{4}\right ]} \left (W_{\left (E\right ) \nu _{1} \nu _{2}}^{\mu _{1} \mu _{2}} W_{\left (E\right ) \nu _{3} \nu _{4}}^{\mu _{3} \mu _{4}} +2 W_{\left (E\right ) \nu _{1} \nu _{2}}^{\mu _{1} \mu _{2}} W_{\left (N E\right ) \nu _{3} \nu _{4}}^{\mu _{3} \mu _{4}} +W_{\left (N E\right ) \nu _{1} \nu _{2}}^{\mu _{1} \mu _{2}} W_{\left (N E\right ) \nu _{3} \nu _{4}}^{\mu _{3} \mu _{4}}\right )\text{\ \ }\;\text{.}\;\;
\end{equation}
Note that the renormalized AdS action for Einstein gravity arises as the first term in the above relation if the coupling is chosen as $\alpha _{C G} =\ell ^{2}/64 \pi  G$. Thus, the equivalent form for the CG action is
\begin{equation}I_{C G} =I_{r e n}^{\left (E\right )} -\frac{\ell ^{2}}{16 \pi  G} \gamma  \int _{M}d^{4} x \sqrt{ -g} \delta _{\left [\mu _{1} \mu _{2}\right ]}^{\left [\nu _{1} \nu _{2}\right ]} (G_{\nu _{1}}^{\mu _{1}} -\frac{\gamma }{2} B_{\nu _{1}}^{\mu _{1}}) B_{\nu _{2}}^{\mu _{2}}\; , \label{CGactiondecomp}
\end{equation}  where the Einstein tensor is written as $G_{\mu }^{\nu } = -4\delta _{\left [\mu  \gamma \delta \right ]}^{\left [\nu \alpha \beta   \right ]}W_{\left (E\right ) \alpha \beta }^{\gamma \delta }$  and $\gamma  =\ell ^{2}/4$. 

This formulation for the CG action makes manifest an interesting property: when the
Bach tensor vanishes, Eq.(\ref{CGactiondecomp}) reduces to the action of Einstein AdS theory. This equivalence
is valid not only for Einstein spacetimes but also for conformally Einstein ones, as Bach flatness is a sufficient condition. 

The
emergence of Einstein gravity can be seen at the level of the first variation of the action, as well. When EOM for CG hold, the surface term in Eq.(\ref{confgravar})
becomes
\begin{equation}\delta  I_{C G} =\frac{\ell ^{2}}{64 \pi  G} \int _{ \partial M}d^{3} x \sqrt{ -h} \delta _{\left [\mu _{1} \mu _{2} \mu _{3} \mu _{4}\right ]}^{\left [\nu _{1} \nu _{2} \nu _{3} \nu _{4}\right ]} [n_{\nu _{1}} \delta  \Gamma _{\lambda  \nu _{2}}^{\mu _{1}} g^{\mu _{2} \lambda } \left (W_{\left (E\right ) \nu _{3} \nu _{4}}^{\mu _{3} \mu _{4}} +W_{\left (N E\right ) \nu _{3} \nu _{4}}^{\mu _{3} \mu _{4}}\right ) +n^{\mu _{1}}  \nabla _{\nu _{1}}W_{\left (N E\right ) \nu _{2} \nu _{3}}^{\mu _{2} \mu _{3}} \left (g^{ -1} \delta  g\right )_{\nu _{4}}^{\mu _{4}}]
\end{equation}  for the decomposition of the Weyl tensor mentioned above. Here, the Bianchi identity has been used to get rid of the
$W_{(E)}$ part under the covariant derivative. It is easy to cast the whole surface term in the form
\begin{multline}\delta  I_{C G} =\frac{\ell ^{2}}{64 \pi  G} \int _{ \partial M}d^{3} x \sqrt{ -h} \delta _{\left [\mu _{1} \mu _{2} \mu _{3} \mu _{4}\right ]}^{\left [\nu _{1} \nu _{2} \nu _{3} \nu _{4}\right ]} n_{\nu _{1}} \delta  \Gamma _{\lambda  \nu _{2}}^{\mu _{1}} g^{\mu _{2} \lambda } W_{\left (E\right ) \nu _{3} \nu _{4}}^{\mu _{3} \mu _{4}} - \\
 -\frac{\ell ^{2}}{32 \pi  G} \gamma  \int _{ \partial M}d^{3} x \sqrt{ -h} \delta _{\left [\mu _{1} \mu _{2} \mu _{3}\right ]}^{\left [\nu _{1} \nu _{2} \nu _{3}\right ]} \left [n_{\nu _{1}} \delta  \Gamma _{\lambda  \nu _{2}}^{\mu _{1}} g^{\mu _{2} \lambda } B_{\nu _{3}}^{\mu _{3}} +n^{\mu _{1}}  \nabla _{\nu _{1}}B_{\nu _{2}}^{\mu _{2}} \left (g^{ -1} \delta  g\right )_{\nu _{3}}^{\mu _{3}}\right ] \label{varcgdecomp}\end{multline}  Notice that $\delta  \Gamma $ in the first term may contain non-Einstein contributions from the metric, such that we cannot properly say that this term
matches Eq.(\ref{varadsaction}). However, without loss of generality, switching off non-Einstein modes in the
metric, what implies a vanishing Bach tensor leads to an equivalence between general relativity with cosmological constant and CG, at the level of the variation
of the action. 

\section{ Critical Gravity}
In Ref.\cite{Lu2011}, a new higher-derivative gravity theory
was introduced, defined by a specific point in the space of parameters, which are the couplings of the quadratic-curvature terms added on top of Einstein
gravity. Indeed, critical gravity action
\begin{equation}I_{c r i t i c a l} =\frac{1}{16 \pi  G} \int _{M}d^{4} x \sqrt{ -g} \left [R -2 \Lambda  +\frac{3}{2 \Lambda } \left (R_{\mu  \nu } R^{\mu  \nu } -\frac{1}{3} R^{2}\right )\right ]
\end{equation}  is singled out by the requirement that the massive graviton modes turn massless and the massive scalar fields are absent.

This precise combination of quadratic terms in the curvature appears at the boundary of AdS$_{5}$ spacetimes, as it is proportional to the conformal anomaly of the dual conformal field theory (CFT) living there \cite{Balasubramanian2000}.

It was shown in Ref. \cite{Miskovic2014} that an
equivalent form for critical gravity action is
\begin{equation}I_{c r i t i c a l} =I_{r e n}^{\left (E\right )} -\frac{\ell ^{2}}{256 \pi  G} \int _{M}d^{4} x \sqrt{ -g} \delta _{\left [\mu _{1} \mu _{2} \mu _{3} \mu _{4}\right ]}^{\left [\nu _{1} \nu _{2} \nu _{3} \nu _{4}\right ]} W_{\nu _{1} \nu _{2}}^{\mu _{1} \mu _{2}} W_{\nu _{3} \nu _{4}}^{\mu _{3} \mu _{4}}\; , \label{CriticalGrav}
\end{equation}  

This casts the critical gravity action (\ref{CGactiondecomp}) into the form
\begin{equation}I_{c r i t i c a l} = -\frac{\ell ^{4}}{64\pi G}\int _{M}d^{4} x \sqrt{ -g} \delta _{\left [\mu _{1} \mu _{2}\right ]}^{\left [\nu _{1} \nu _{2}\right ]} \left ( \frac{\ell ^{2}}{8} B_{\nu _{1}}^{\mu _{1}} - G_{\nu _{1}}^{\mu _{1}}\right ) B_{\nu _{2}}^{\mu _{2}}\text{\thinspace \ }\;\text{.}\;\;
\end{equation}  It is then straightforward to show that the on-shell action can be written down as
a quadratic contribution in the Bach tensor
\begin{equation}I_{c r i t i c a l} =\frac{\ell ^{6}}{512 \pi  G} \int _{M}d^{4} x \sqrt{ -g} \delta _{\left [\mu _{1} \mu _{2}\right ]}^{\left [\nu _{1} \nu _{2}\right ]} B_{\nu _{1}}^{\mu _{1}} B_{\nu _{2}}^{\mu _{2}}\text{\thinspace \ }\;\text{.}\;\;
\end{equation}  The non-Einstein part of the curvature (Bach tensor) is the only one to survive around the critical point in critical gravity. This provides a further explanation of the fact that the action for critical gravity is 0 for any Einstein space (so does the black hole mass and the entropy as stated in the original reference Ref.\cite{Lu2011}.

\section{ Conclusions}
In this paper, we have provided a generic argument on the equivalence between Einstein gravity with a cosmological constant and conformal gravity
for Bach-flat spacetimes, for a fixed value of the CG coupling constant. Upon switching on the trace-free part of the Ricci tensor, we assume a splitting
of the Weyl tensor that is suitable for a broader class of spaces, beyond Einstein gravity, which plays a crucial role in the derivation presented. 

The transition from CG to Einstein gravity is also useful to identify the non-Einstein modes in the critical gravity action as proportional to
the square of the Bach tensor. This result is consistent with a vanishing mass for both black holes and graviton excitations in the Einstein branch of the
theory around the critical point. 

The formulation presented here is expected to provide a shortcut in the computation of holographic
correlation functions in critical gravity \cite{Johansson2012,Naseh}. Indeed,
using an asymptotic expansion of the metric that takes into account new boundary sources associated to higher-derivative field equations allows us to identify
the appropriate counterterms that render the action finite for logarithmic modes, and to extract the physical information on the boundary CFT \cite{AnastasiouOlea}.

\begin{acknowledgments}The authors thank G. Giribet, O. Miskovic, P. Sundell, and T. Zojer for interesting discussions. G.A. is supported by a Universidad Andres Bello (UNAB) PhD scholarship.
The work of R.O. is funded in part by Fondo Nacional de Desarollo Cientifico y Tecnologico (FONDECYT) Grant No. 1131075, UNAB Grant No. DI-1336-16/R, and Comision Nacional de Investigacion Cientifica y Tecnologica (CONICYT) Grant No. DPI 20140115.

\end{acknowledgments}
\end{document}